\documentclass[twocolumn,superscriptaddress,preprintnumbers,amsmath,amssymb,prd]{revtex4}
\usepackage{graphicx}

\begin{document}
\title{Casimir entropy and nonlocal response functions to the off-shell
quantum fluctuations}

\author{
G.~L.~Klimchitskaya}
\affiliation{Central Astronomical Observatory at Pulkovo of the
Russian Academy of Sciences, Saint Petersburg,
196140, Russia}
\affiliation{Institute of Physics, Nanotechnology and
Telecommunications, Peter the Great Saint Petersburg
Polytechnic University, Saint Petersburg, 195251, Russia}

\author{
V.~M.~Mostepanenko}
\affiliation{Central Astronomical Observatory at Pulkovo of the
Russian Academy of Sciences, Saint Petersburg,
196140, Russia}
\affiliation{Institute of Physics, Nanotechnology and
Telecommunications, Peter the Great Saint Petersburg
Polytechnic University, Saint Petersburg, 195251, Russia}
\affiliation{Kazan Federal University, Kazan, 420008, Russia}

\begin{abstract}
The nonlocal response functions to quantum fluctuations
are used to find asymptotic expressions for the Casimir free
energy and entropy at arbitrarily low temperature in the configuration
of two parallel metallic plates. It is shown that by introducing an
alternative nonlocal response to the off-the-mass-shell fluctuations the
Lifshitz theory is brought into agreement with the requirements of
thermodynamics. According to our results, the Casimir entropy
calculated using the nonlocal response functions, which take into
account dissipation of conduction electrons, remains positive
and monotonously goes to zero with vanishing temperature, i.e.,
satisfies the Nernst heat theorem. This is true for both plates
with perfect crystal lattices and for lattices with defects of
structure. The obtained results are discussed in the context of the
Casimir puzzle.
\end{abstract}

\maketitle

\newcommand{\ri}{{\rm i}}
\newcommand{\il}{{{\rm i}\xi_l}}
\newcommand{\kb}{{k_{\bot}}}
\newcommand{\kk}{{({\rm i}\xi_l,k_{\bot})}}
\newcommand{\kTl}{{k_l^T}}
\newcommand{\vk}{\mbox{\boldmath$k$}}
\newcommand{\eps}{{\varepsilon}}
\newcommand{\teps}{{\tilde{\varepsilon}}}
\newcommand{\teT}{{\tilde{\varepsilon}}^{\rm Tr}}
\newcommand{\teL}{{\tilde{\varepsilon}}^{\rm L}}
\newcommand{\teTD}{{\tilde{\varepsilon}}_{D}^{\rm Tr}}
\newcommand{\teLD}{{\tilde{\varepsilon}}_{D}^{\rm L}}
\newcommand{\elTD}{{\tilde{\varepsilon}}_{D,l}^{\rm Tr}}
\newcommand{\elLD}{{\tilde{\varepsilon}}_{D,l}^{\rm L}}
\newcommand{\rTM}{{r_{\rm TM}}}
\newcommand{\rTE}{{r_{\rm TE}}}
\newcommand{\rA}{{r_{\alpha}}}
\newcommand{\aT}{(a,T)}
\newcommand{\vt}{v^{\rm Tr}}
\newcommand{\vl}{v^{\rm L}}
\newcommand{\tvt}{{\tilde{v}^{\rm Tr}}}
\newcommand{\tvl}{{\tilde{v}^{\rm L}}}
\newcommand{\bt}{{\tilde{b}}}
\newcommand{\gt}{{\tilde{\gamma}_0}}
\newcommand{\cFa}{{\cal{F}_{\alpha}}(a,T)}
\newcommand{\acF}{{\cal{F}_{\alpha}}}
\newcommand{\ocF}{{\cal{F}}}
\newcommand{\dT}{{\Delta_T}}
\newcommand{\zy}{{({\rm i}\zeta_l,y,T)}}
\newcommand{\ozy}{{({\rm i}\zeta_l,y,0)}}
\newcommand{\oyt}{{(0,y,T)}}
\newcommand{\daT}{{\Delta_T^{\rm impl}}}
\newcommand{\deT}{{\Delta_{T}^{\rm expl}}}
\newcommand{\dbT}{{\Delta_{T\!,\,l=0}^{\rm expl}\,}}
\newcommand{\dcT}{{\Delta_{T\!,\,l\geqslant 1}^{\rm expl}\,}}
\section{Introduction}

The Casimir effect is the most impressive physical phenomenon demonstrating
an existence of the zero-point oscillations of quantum fields. As was shown
by H. B. G. Casimir \cite{1}, two parallel ideal metal plates at zero
temperature placed in vacuum at a distance $a$ are attracted by the force
per unit area $-\pi^2\hbar c/(240a^4)$, which is completely determined by
the zero-point oscillations of the electromagnetic field. Shortly thereafter
E.M. Lifshitz created unified theory of the van der Waals and Casimir
forces between two parallel plates made of any materials kept at temperature
$T$ in thermal equilibrium with the environment \cite{2,4}. In the
framework of this theory, the Casimir force is caused by the joint action
of zero-point and thermal fluctuations. In so doing the response of plate
materials to quantum fluctuations is described by the frequency-dependent
dielectric permittivities, whereas the van der Waals force proves to be
a special case of the Casimir force when separations between the plates
are so small that the speed of light $c$ can be considered as infinitely
large. In recent years, the Lifshitz theory was generalized for the case
of two compact arbitrarily shaped bodies \cite{5,6,7}. As a result, it has
become possible to express the Casimir force between any two bodies
using the formalism of thermal quantum field theory in terms of the
reflection amplitudes of quantum fluctuations on their boundary surfaces.

In the early twenty first century, precise measurements of the Casimir
interaction between metallic surfaces performed by R. S. Decca et al.
\cite{8,9,10,11} discovered a serious discrepancy between theoretical
Casimir forces calculated using the Lifshitz theory and the measurement
data. In calculations, the response of metal to low-frequency quantum
fluctuations was described by the dissipative Drude model, i.e., in the
same manner as to real electromagnetic fields on the mass shell with nonzero
field strengths. The surprising thing is that the Lifshitz theory agreed
with the same experimental data if the dissipationless plasma response
function was used which should not be applicable at low frequencies.
More recently, the results of Refs.~\cite{8,9,10,11} were conclusively confirmed by
many experiments \cite{12,13,14,15,16,17,18,19,19a} (see also reviews
\cite{20,21}). This situation has been characterized in the literature
as the Casimir puzzle or Casimir conundrum \cite{22,23,24,25,25a}.

At the same time it was shown that the Casimir entropy calculated within
the Lifshitz theory for metals with perfect crystal lattices goes to
a negative quantity depending on the plate parameters with vanishing
temperature, i.e., violates the third law of thermodynamics, the Nernst
heat theorem, if the Drude response function to quantum fluctuations is
used \cite{26,27,28,29,30}. What is more, employing the experimentally
consistent plasma model brings the Lifshitz theory in agreement with
thermodynamics \cite{26,27,28,29,30}. The reasons why the evidently
inapplicable at low frequencies response function leads to so good
results whereas the well tested Drude response fails to reach an
agreement with the measurement data and thermodynamical laws remained
unknown.

It should be mentioned that several authors argued for an accord
between the Lifshitz theory employing the Drude response function and
thermodynamics \cite{31,32,33}. In favor of this statement they have
shown that if the crystal lattice of a metal contains some fraction of
defects the Casimir entropy abruptly jumps to zero at very low
temperature starting from negative values. For the perfect crystal
lattice, however, it remained impossible to reconcile the Casimir
entropy calculated using the Drude model with thermodynamics. Besides,
the consideration of imperfect crystal lattices was not helpful to
attain an agreement between the experimental data and theoretical
predictions. Many attacks to this problem have been undertaken in
the literature (see, e.g.,
Refs.~\cite{34,35,36,37,38,39,40,41,42,43,44,45,46,47})
but no wholly satisfactory solution is yet available.

Quite recently a novel approach to the resolution of the Casimir
puzzle has been proposed \cite{48}. This approach assumes that the
electromagnetic response of a metal to quantum fluctuations is spatially
nonlocal. However, unlike the commonly applied nonlocal response
functions \cite{35,49,50,51,52,54,55}, the suggested res\-ponse leads
to nearly the same results, as the standard Drude function, for the
quantum fluctuations on the mass shell, but to significantly different
results for the off-shell fluctuations. The proposed alternative
Drude-like transverse and longitudinal dielectric permittivities
take a complete account of dissipation at low frequencies, as does
the conventional Drude model. It was shown, however, that the Lifshitz
theory using the nonlocal response functions to quantum fluctuations
predicts values of the Casimir force which agree with the
measurement data as well as those predicted using the dissipationless
plasma model. It was also shown that the suggested nonlocal
dielectric functions describe correctly the response of metal to
real electromagnetic fields on the mass shell in optical experiments
and obey the Kramers-Kronig relations. In essence, the possibility to
introduce the nonlocal Drude-like functions is based on
the fact that the electromagnetic response of a metal to quantum
fluctuations off the mass shell is not directly measurable.
Experimentally, some indirect information can be obtained only
concerning the longitudinal dielectric function \cite{55} which,
however, does not affect a comparison between
theoretical Casimir forces and the measurement data \cite{48}.

In this paper, we find the Casimir free energy and entropy given
by the Lifshitz theory and the nonlocal response
functions introduced in Ref.~\cite{48}. We derive an analytic asymptotic
expression for the Casimir free energy at arbitrarily low
temperature and show that for metals with perfect crystal lattices
the respective Casimir entropy is positive and goes to zero with
vanishing temperature. Thus, as opposed to the conventional Drude
model, the Lifshitz theory using nonlocal response functions
is in agreement with the laws of thermodynamics. It is shown also
that this result remains valid for crystal lattices with some fraction
of defects, i.e., the Casimir entropy remains positive and monotonously
goes to zero with decreasing temperature according to the Nernst heat
theorem.

The structure of the paper is as follows. In Sec.~II, we briefly
introduce the nonlocal response functions and basic
expressions of the Lifshitz theory for the Casimir free energy
in spatially nonlocal case. Section~III contains derivation
of the low-temperature expansion for the Casimir free energy
and entropy using these response functions for metals
with perfect crystal lattices. In Sec.~IV, the case of crystal
lattices with defects of structure is considered. Section~V is for our
conclusions and a discussion. The Appendix A contains some
details of mathematical derivations.

\section{The nonlocal response functions to quantum
fluctuations and the Casimir free energy}

We consider two parallel thick metallic plates at temperature $T$
in thermal equilibrium with the environment. The separation distance
between plates is notated $a$. For good metals (Au, for instance)
the plates of more than 100~nm thickness can be considered as
semispaces when calculating the Casimir free energy and pressure
\cite{20}. According to the Lifshitz theory, the free energy of the
Casimir interaction per unit area of the plates can be presented in
the form \cite{2,4} (see also Refs.~\cite{20,21} for the current notations)
\begin{equation}
{\cal F}\aT=\sum_{\alpha}{\cal F}_{\alpha}\aT,
\label{eq1}
\end{equation}
\noindent
where the sum is over two independent polarizations of the
electromagnetic field, transverse magnetic ($\alpha$ = TM) and
transverse electric ($\alpha$ = TE), and
\begin{eqnarray}
&&
{\cal F}_{\alpha}\aT=\frac{k_BT}{2\pi}\sum_{l=0}^{\infty}{\vphantom{\sum}}^{\prime}
\int_0^{\infty}\!\!\!\kb d\kb
\nonumber \\
&&~~~~~~~~~~~~~~\times
\ln\left[1-r_{\alpha}^2(\il,\kb,T)e^{-2aq_l}\right].
\label{eq2}
\end{eqnarray}
\noindent
Here, $k_B$ in the Boltzmann constant,
the prime on the summation sign in $l$ divides the term with $l=0$ by 2,
$\kb$ is the magnitude of the
projection of wave vector {\vk} on the plane of plates
(this plane is perpendicular to the Casimir force),
the Matsubara frequencies are $\xi_l=2\pi k_BTl/\hbar$ with
$l=0,\,1,\,2,\,\ldots$, and
$q_l=q_l(k_{\bot})=(k_{\bot}^2+\xi_l^2/c^2)^{1/2}$.

The quantities $\rA$ in Eq.~(\ref{eq2}) have a meaning of the reflection
coefficients calculated at the pure imaginary Matsubara frequencies.
In the spatially local case they coincide with the familiar Fresnel
coefficients
\begin{eqnarray}
&&
\rTM(\il,\kb,T)=\frac{\eps_lq_l-k_l}{\eps_lq_l+k_l},
\nonumber \\
&&
\rTE(\il,\kb,T)=\frac{q_l-k_l}{q_l+k_l},
\label{eq3}
\end{eqnarray}
\noindent
where $k_l=k_l(k_{\bot},T)=(k_{\bot}^2+\eps_l\xi_l^2/c^2)^{1/2}$
and $\eps_l=\eps(\il,T)$
is the dielectric permittivity  which describes the local response of a metal
to quantum fluctuations. Our notations underline that the dielectric permittivity
of metals explicitly depends on $T$ through the relaxation parameter
(see below). The reflection coefficients (\ref{eq3}) also possess an implicit
dependence on $T$ through the Matsubara frequencies.

In our case the dielectric response  of a metal is spatially nonlocal and, thus,
is described by two independent permittivities, the transverse one,
$\eps^{\rm Tr}(\omega,\vk,T)$,
and the longitudinal one, $\eps^{\rm L}(\omega,\vk,T)$, where the wave vector
$\vk=(\kb,k^3)$ \cite{55,56}. We recall that $\eps^{\rm Tr}$ and $\eps^{\rm L}$
describe the response of a metal to the transverse electric field which is
perpendicular to $\vk$ and to the longitudinal one which is parallel to $\vk$,
respectively. As was argued in Ref.~\cite{48}, in the plane-parallel Casimir geometries
the nonlocal permittivities should depend only on $\kb$. In this case it was
proven \cite{48} that the reflection coefficients in Eq.~(\ref{eq2}) are given by
\begin{eqnarray}
&&
\rTM(\il,\kb,T)=\frac{\eps_{l}^{\rm Tr}q_l-k_l^{\rm Tr}-\kb\left(\eps_l^{\rm Tr}-
\eps_l^{\rm L}\right)\left({\eps_l^{\rm L}}\right)^{-1}}{\eps_{l}^{\rm Tr}
q_l+k_l^{\rm Tr}+\kb\left(\eps_l^{\rm Tr}-
\eps_l^{\rm L}\right)\left({\eps_l^{\rm L}}\right)^{-1}},
\nonumber \\
&&
\rTE(\il,\kb,T)=\frac{q_l-k_l^{\rm Tr}}{q_l+k_l^{\rm Tr}},
\label{eq4}
\end{eqnarray}
\noindent
where $\eps_{l}^{\rm Tr}=\eps^{\rm Tr}(\il,\kb,T)$,
$\eps_{l}^{\rm L}=\eps^{\rm L}(\il,\kb,T)$ and
\begin{equation}
k_{l}^{\rm Tr}=k_l^{\rm Tr}(\kb,T)=
\left[k_{\bot}^2+\eps_{l}^{\rm Tr}\frac{\xi_l^2}{c^2}\right]^{1/2}.
\label{eq5}
\end{equation}

The alternative Drude-like nonlocal response functions suggested in Ref.~\cite{48}
take the form
\begin{eqnarray}
&&
\teTD(\omega,\kb,T)=1-\frac{\omega_p^2}{\omega[\omega+\ri\gamma(T)]}
\left(1+\ri\frac{\vt\kb}{\omega}\right),
\nonumber \\
&&
\teLD(\omega,\kb,T)=1-\frac{\omega_p^2}{\omega[\omega+\ri\gamma(T)]}
\left(1+\ri\frac{\vl\kb}{\omega}\right)^{-1}\!\!\!,
\label{eq6}
\end{eqnarray}
\noindent
where $\omega_p$ is the plasma frequency, $\gamma(T)$ is the relaxation
parameter, and $\vt,\,\,\vl$ are the constants of the order of Fermi
velocity $v_F\sim 0.01 c$. In the local limit $\kb=0$ and the permittivities
(\ref{eq6}) reduce to the standard permittivity of the Drude model
\begin{eqnarray}
&&
\teTD(\omega,0,T)=\teLD(\omega,0,T)=\eps_D(\omega,T)
\nonumber \\
&&~~~~~~~~~~~~
=
1-\frac{\omega_p^2}{\omega[\omega+\ri\gamma(T)]}.
\label{eq7}
\end{eqnarray}

For the electromagnetic fields on the mass shell it holds
$\kb\leqslant\omega/c$. As a result one obtains
\begin{equation}
\frac{v^{\rm Tr, L}\kb}{\omega}\sim \frac{v_F}{c}\,\frac{c\kb}{\omega}
\leqslant\frac{v_F}{c}\ll 1.
\label{eq8}
\end{equation}
\noindent
Because of this, for the on-shell fields the dielectric permittivities
(\ref{eq6}) lead to nearly the same results as the commonly used local
Drude permittivity (\ref{eq7}). As to the off-shell fluctuations, the
quantity $v^{\rm Tr, L}\kb/\omega$ can be of the order of and even larger
than unity depending on the value of $\kb$. Thus, by using the
permittivities (\ref{eq6}) at the pure imaginary Matsubara frequencies
\begin{eqnarray}
&&
\elTD=1+\frac{\omega_p^2}{\xi_l[\xi_l+\gamma(T)]}
\left(1+\frac{\vt\kb}{\xi_l}\right),
\nonumber \\
&&
\elLD=1+\frac{\omega_p^2}{\xi_l[\xi_l+\gamma(T)]}
\left(1+\frac{\vl\kb}{\xi_l}\right)^{-1}
\label{eq9}
\end{eqnarray}
\noindent
one can restore an agreement between theoretical predictions of the Lifshitz
theory with taken into account dissipation properties of conduction electrons and
the measurement data \cite{48}.

In the next section, we consider the asymptotic behavior of the Casimir free
energy (\ref{eq1}), (\ref{eq2}) with the reflection coefficients (\ref{eq4}) and
dielectric permittivities (\ref{eq9}) at arbitrarily low temperature. For this purpose
it is convenient to present both contributions to the free energy (\ref{eq1}) as the
sum of the zero-temperature terms and the thermal corrections to them
\begin{equation}
\cFa=E_{\alpha}(a)+\dT\cFa.
\label{eq10}
\end{equation}

The quantity $E_{\alpha}(a)$ is obtained from $\ocF_{\alpha}$ defined in Eq.~(\ref{eq2})
by putting $T=0$ and replacing the discrete Matsubara frequencies $\xi_l$ with
a continuous variable $\xi$. In so doing the sum in $l$ is replaced with an integral
\begin{equation}
k_BT\sum_{l=0}^{\infty}{\vphantom{\sum}}^{\prime}\to
\frac{\hbar}{2\pi}\int_0^{\infty}\!\!d\xi,
\label{eq11}
\end{equation}
\noindent
and one obtains
\begin{eqnarray}
&&
E_{\alpha}(a)=\frac{\hbar}{4\pi^2}\int_0^{\infty}\!\!\!d\xi
\int_0^{\infty}\!\!\!\kb d\kb
\nonumber \\
&&~~~~~~~~~~~~~~\times
\ln\left[1-r_{\alpha}^2(\ri\xi,\kb,0)e^{-2aq}\right].
\label{eq12}
\end{eqnarray}
\noindent
Here, the reflection coefficients are given by
\begin{eqnarray}
&&
\rTM(\ri\xi,\kb,0)
\nonumber\\
&&~~
=\frac{\eps^{\rm Tr(0)}q-k^{\rm Tr(0)}-\kb\left(\eps^{\rm Tr(0)}-
\eps^{\rm L(0)}\right)\left({\eps^{\rm L(0)}}\right)^{-1}}{\eps^{\rm Tr(0)}q+
k^{\rm Tr(0)}+\kb\left(\eps^{\rm Tr(0)}-
\eps^{\rm L(0)}\right)\left({\eps^{\rm L(0)}}\right)^{-1}},
\nonumber \\
&&
\rTE(\ri\xi,\kb,0)=\frac{q-k^{\rm Tr(0)}}{q+k^{\rm Tr(0)}},
\label{eq13}
\end{eqnarray}
\noindent
where $\eps^{\rm Tr(0)}=\eps^{\rm Tr}(\ri\xi,\kb,0)$,
$\eps^{\rm L(0)}=\eps^{\rm L}(\ri\xi,\kb,0)$ and
\begin{eqnarray}
&&
q= q(\kb)=
\left[k_{\bot}^2+\eps^{\rm Tr(0)}\frac{\xi^2}{c^2}\right]^{1/2},
\nonumber \\
&&
k^{\rm Tr(0)}= k^{\rm Tr}(\kb,0)=
\left[k_{\bot}^2+\eps^{\rm Tr(0)}\frac{\xi^2}{c^2}\right]^{1/2}.
\label{eq14}
\end{eqnarray}

Equation (\ref{eq10}) can be considered as a definition of the thermal
correction which presents a simple and stra\-ightforward way for its calculation
in the limiting case of low temperature.

\section{Low-temperature behavior of the Casimir free energy  and entropy using the
nonlocal response functions  for perfect  crystal lattices}

Now we consider asymptotic behavior of the thermal correction $\dT\acF$ defined
in Eq.~(\ref{eq10}) at low temperature where the Casimir free energy $\acF$ and
energy $E_{\alpha}$ are defined in Eqs.~(\ref{eq2}) and (\ref{eq12}), respectively.
It is convenient to perform the asymptotic expansion using the dimensionless
variables
\begin{equation}
\zeta_l=\frac{2a\xi_l}{c}, \qquad
y=2aq_l.
\label{eq15}
\end{equation}

In terms of these variables the thermal correction takes the form
\begin{eqnarray}
&&
\dT\acF\aT=\frac{k_BT}{8\pi a^2}\!\sum_{l=0}^{\infty}{\vphantom{\sum}}^{\prime}
\!\!\int_{\zeta_l}^{\infty}\!\!\!\!\!y dy
\ln\left[1-r_{\alpha}^2(\ri\zeta_l,y,T)e^{-y}\right]
\nonumber \\
&&~
-\frac{\hbar c}{32\pi^2a^3}\int_0^{\infty}\!\!\!d\zeta
\int_{\zeta}^{\infty}\!\!\!y dy
%\nonumber \\
%&&~~~~~~~~~~~~~~\times
\ln\left[1-r_{\alpha}^2(\ri\zeta,y,0)e^{-y}\right],
\label{eq16}
\end{eqnarray}
\noindent
where $\zeta=2a\xi/c$ similar to (\ref{eq15}).

We introduce in the second line of Eq.~(\ref{eq16}) the integration variable
$t=\zeta/\tau$, where $\tau=4\pi k_BTa/(\hbar c)$, and add and subtract
on the right-hand side of Eq.~(\ref{eq16}) the following quantity:
\begin{equation}
\frac{k_BT}{8\pi a^2}\sum_{l=0}^{\infty}{\vphantom{\sum}}^{\prime}
\!\int_{\zeta_l}^{\infty}\!\!\!\!y dy
\ln\left[1-r_{\alpha}^2(\ri\zeta_l,y,0)e^{-y}\right].
\label{eq17}
\end{equation}
\noindent
Then, taking into account, that $\zeta_l=\tau l$, one can identically rewrite
Eq.~(\ref{eq16}) as
\begin{equation}
\dT\acF\aT=\deT\acF\aT+\daT\acF\aT,
\label{eq18}
\end{equation}
\noindent
\vspace*{-3mm}
where
\begin{equation}
\deT\acF\aT=\frac{k_BT}{8\pi a^2}\!\sum_{l=0}^{\infty}{\vphantom{\sum}}^{\prime}
\!\!\int_{\tau l}^{\infty}\!\!\!\!\!y dy
\ln\frac{1-r_{\alpha}^2(\ri\tau l,y,T)e^{-y}}{1-r_{\alpha}^2(\ri\tau l,y,0)e^{-y}}
\label{eq19}
\end{equation}
\noindent
\vspace*{-3mm}
and
\begin{equation}
\daT\acF\aT=\frac{k_BT}{8\pi a^2}\left[\sum_{l=0}^{\infty}{\vphantom{\sum}}^{\prime}
\Phi_{\alpha}(\tau l)-\int_0^{\infty}\!\!\!dt\, \Phi_{\alpha}(\tau t)\right],
\label{eq20}
\end{equation}
\noindent
where
\begin{equation}
\Phi_{\alpha}(x)\equiv \int_x^{\infty}\!\!\!ydy\ln\left[1-r_{\alpha}^2(\ri x,y,0)
e^{-y}\right].
\label{eq21}
\end{equation}

The quantity $\deT\ocF$ is called the explicit thermal correction. It vanishes for
the response functions and hence for the reflection coefficients which do not
possess an explicit dependence on temperature as a parameter.
As to the quantity $\daT\ocF$, which is called the implicit thermal correction,
it depends on temperature only through the Matsubara frequencies.

We start with the implicit thermal correction  which, as shown below,  provides
the dominant contribution to the low-temperature dependence of the Casimir free
energy. Using the Abel-Plana formula for the difference between the sum and the
integral \cite{57}, one can identically rearrange Eqs.~(1) and (20) to
\begin{eqnarray}
&&
\daT\ocF\aT=\frac{\ri k_BT}{8\pi a^2}\int_0^{\infty}\frac{dt}{e^{2\pi t}-1}
\nonumber \\
&&~~~~~~~~~~~\times
\sum_{\alpha}\left[
\Phi_{\alpha}(\ri\tau t)-\Phi_{\alpha}(-\ri\tau t)\right].
\label{eq22}
\end{eqnarray}

In subsequent derivations one should take into account the dependence of the
relaxation parameter $\gamma$ on $T$. In this section, we consider metals with
perfect crystal lattices. In this case, the relation $\gamma(T)=bT^2$, where
$b$ is some constant coefficient, is followed starting from the liquid helium
temperature down to zero temperature owing to the electron-electron scattering
\cite{58}. Rewriting Eq.~(\ref{eq9}) at $T=0$ (where $\xi_l$ is replaced with $\xi$)
in terms of the dimensionless variables introduced above, one obtains
\begin{eqnarray}
&&
\teTD{\vphantom{\teps}}^{(0)}=\teTD(\ri x,y,0)
\nonumber \\
&&~~~~
=1+\frac{\tilde{\omega}_p^2}{x^2}
\left(1+{\tvt}\frac{\sqrt{y^2-x^2}}{x}\right),
\nonumber \\
&&
\teLD{\vphantom{\teps}}^{(0)}=\teLD(\ri x,y,0)
\nonumber \\
&&~~~~
=1+\frac{\tilde{\omega}_p^2}{x^2}
\left(1+{\tvl}\frac{\sqrt{y^2-x^2}}{x}\right)^{-1}\!\!\!,
\label{eq23}
\end{eqnarray}
\noindent
where $\tilde{\omega}_p=2a\omega_p/c$, $\tvt=\vt/c$, and $\tvl=\vl/c$ are the
dimensionless plasma frequency and respective velocities.

As  a result, the reflection coefficients $\rA(\ri x,y,0)$, entering Eq.~(\ref{eq21}),
take the form
\begin{widetext}
\begin{eqnarray}
&&
\rTM(\ri x,y,0)
\nonumber
=\frac{\teTD{\vphantom{\teps}}^{(0)}y-\sqrt{y^2+(\teTD{\vphantom{\teps}}^{(0)}-1)x^2}-
\left[\teTD{\vphantom{\teps}}^{(0)}-\teLD{\vphantom{\teps}}^{(0)}
\right]\left(\teLD{\vphantom{\teps}}^{(0)}\right)^{-1}\sqrt{y^2-x^2}}{\teTD{\vphantom{\teps}}^{(0)}y+
\sqrt{y^2+(\teTD{\vphantom{\teps}}^{(0)}-1)x^2}+
\left[\teTD{\vphantom{\teps}}^{(0)}-\teLD{\vphantom{\teps}}^{(0)}
\right]\left(\teLD{\vphantom{\teps}}^{(0)}\right)^{-1}\sqrt{y^2-x^2}},
\nonumber \\
&&
\rTE(\ri x,y,0)=\frac{y-\sqrt{y^2+(\teTD{\vphantom{\teps}}^{(0)}-1)x^2}}{y+
\sqrt{y^2+(\teTD{\vphantom{\teps}}^{(0)}-1)x^2}}.
\label{eq24}
\end{eqnarray}
\end{widetext}
\noindent
{}From Eqs.~(\ref{eq23}) and (\ref{eq24}) it is easily seen that in the limiting case
of zero temperature, i.e., for  $x=\tau t\to 0$, it holds
\begin{equation}
\lim_{x\to 0}\rTM(\ri x,y,0)=1, \quad
\lim_{x\to 0}\rTE(\ri x,y,0)=-1.
\label{eq25}
\end{equation}

Now we substitute Eqs.~(\ref{eq23}) and (\ref{eq24}) in Eq.~(\ref{eq21}) and find the first
expansion terms in the powers of small $x$
\begin{eqnarray}
&&
\Phi_{\rm TM}(x)=\int_{0}^{\infty}\!\!\!dy\,y\ln(1-e^{-y})
\nonumber \\
&&~~~~~~
+\frac{4\tvl}{\tilde{\omega}_p^2}x\int_{0}^{\infty}\!\!\!dy
\frac{y^2}{e^y-1}+O(x^2\ln x),
\nonumber \\
&&
\Phi_{\rm TE}(x)=\int_{0}^{\infty}\!\!\!dy\,y\ln(1-e^{-y})
\nonumber \\
&&~~~~~~
+\frac{4}{\tilde{\omega}_p\sqrt{\tvt}}\sqrt{x}\int_{0}^{\infty}\!\!\!dy
\frac{y^{3/2}}{e^y-1}+O(x).
\label{eq26}
\end{eqnarray}
\noindent
{}From Eq.~(\ref{eq26}) it is seen that the leading contribution to the thermal
correction at vanishing temperature is given by the TE mode
\begin{eqnarray}
&&
\Phi(\ri\tau t)-\Phi(-\ri\tau t)=\Phi_{\rm TE}(\ri\tau t)-
\Phi_{\rm TE}(-\ri\tau t)
\nonumber \\
&&~~~~
=\frac{3\ri}{\tilde{\omega}_p\sqrt{\tvt}}\sqrt{2\pi\tau t}
\,\zeta\!\left(\frac{5}{2}\right)+O(\tau)
\nonumber \\
&&~~~~
=\frac{3\ri\pi c}{{\omega}_p}\sqrt{\frac{2k_BT t}{a\vt\hbar}}
\,\zeta\!\left(\frac{5}{2}\right)+O(T),
\label{eq27}
\end{eqnarray}
\noindent
where $\zeta(z)$ is the Riemann zeta function.

Substituting Eq.~(\ref{eq27}) in Eq.~(\ref{eq22}) and integrating with respect to $t$, one
arrives at
\begin{equation}
\daT\ocF(a,T)=-\frac{3c\zeta(3/2)\zeta(5/2)}{32\pi\omega_pa^{5/2}\sqrt{\tvt\hbar}}
(k_BT)^{3/2}+O(T^2).
\label{eq28}
\end{equation}

Now we consider the explicit thermal correction defined in Eq.~(\ref{eq19}). In terms of
dimensionless variables, the dielectric permittivities entering the reflection
coefficients $\rA(\ri\tau l,y,T)$ are obtained from Eq.~(\ref{eq9})
\begin{eqnarray}
&&
\teps_{D,l}^{\rm Tr}=\teTD(\ri\tau l,y,T)
\nonumber \\
&&~~~
=1+\frac{\tilde{\omega}_p^2}{\tau l(\tau l+\bt T^2)}
\left(1+{\tvt}\frac{\sqrt{y^2-\tau^2l^2}}{\tau l}\right),
\nonumber \\
&&
\teps_{D,l}^{\rm L}=\teLD(\ri\tau l,y,T)
\nonumber \\
&&~~
=1+\frac{\tilde{\omega}_p^2}{\tau l(\tau l+\bt T^2)}
\left(1+{\tvl}\frac{\sqrt{y^2-\tau^2l^2}}{\tau l}\right)^{\!-1}\!\!\!\!.
\label{eq29}
\end{eqnarray}
\noindent
Here, the dimensionless relaxation parameter $\tilde{\gamma}(T)=2a\gamma(T)/c=\bt T^2$.
As to the permitti\-vities entering the reflection coefficients $\rA(\ri\tau l,y,0)$,
they are obtained from Eq.~(\ref{eq29}) by putting $T=0$.

To calculate the quantity (\ref{eq19}) at low temperature, we present the reflection
coefficients $\rA(\ri\tau l,y,T)$ as the zero-temperature contributions and the
thermal corrections to them
\begin{equation}
\rA(\ri\tau l,y,T)=\rA(\ri\tau l,y,0)+\dT\rA(\ri\tau l,y,T).
\label{eq30}
\end{equation}
\noindent
It is evident that the thermal corrections $\dT\rA$ go to zero with vanishing
temperature.

We substitute Eq.~(\ref{eq30}) in Eq.~(\ref{eq19}) and expand Eq.~(\ref{eq19}) up to
the first order in the small parameter
\begin{equation}
\frac{\dT\rA(\ri\tau l,y,T)}{\rA(\ri\tau l,y,0)}\ll 1
\label{eq31}
\end{equation}
\noindent
like this was done in the literature for the case of graphene \cite{59}.
The result is
\begin{eqnarray}
&&
\deT\cFa=-\frac{k_BT}{4\pi a^2}\sum_{l=0}^{\infty}{\vphantom{\sum}}^{\prime}
\!\int_{\tau l}^{\infty}\!\!\!\!ydy\rA(\ri\tau l,y,0)
\nonumber \\
&&~~~~~~~~~\times
\frac{\dT\rA(\ri\tau l,y,T)}{e^{y}-\rA^{\! 2}(\ri\tau l,y,0)}.
\label{eq32}
\end{eqnarray}

It is convenient to consider separately the term of Eq.~(\ref{eq32}) with $l=0$,
$\dbT\acF$, and the sum of all terms with $l\geqslant 1$, $\dcT\acF$.
The explicit form of the reflection coefficients $\rA(\ri\tau l,y,T)$
in terms of the dimensionless variables is given by Eq.~(\ref{eq24}) where $x$
should be replaced with $\tau l$ and $\teTD{\vphantom{\teps}}^{(0)}$,
$\teLD{\vphantom{\teps}}^{(0)}$ with $\teps_{D,l}^{\rm Tr}$,
$\teps_{D,l}^{\rm L}$ defined in Eq.~(\ref{eq29}).
As a result, for $\alpha={\rm TM}$, $l=0$ one obtains
\begin{equation}
\rTM(0,y,T)=1-\frac{2\bt T^2\tvl y}{\tilde{\omega}_p^2+2\bt T^2\tvl y}.
\label{eq33}
\end{equation}

Expanding this quantity up to the first order in a small parameter
\begin{equation}
\beta(T)=\frac{\bt T^2\tvl}{\tilde{\omega}_p^2}\ll 1
\label{eq34}
\end{equation}
\noindent
one arrives at
\begin{equation}
\rTM(0,y,T)=1-2\beta(T)y.
\label{eq35}
\end{equation}
\noindent
{}From Eqs.~(\ref{eq33}) and (\ref{eq35}), taking into account Eq.~(\ref{eq30}), we find
\begin{equation}
\rTM(0,y,0)=1, \quad
\dT\rTM(0,y,T)=-2\beta(T)y.
\label{eq36}
\end{equation}

Finally, substituting Eq.~(\ref{eq36}) in Eq.~(\ref{eq32}), we obtain
\begin{equation}
\dbT\ocF_{\rm TM}(a,T)=\frac{k_BT}{4\pi a^2}\beta(T)\int_{0}^{\infty}
\!\!dy\frac{y^2}{e^y-1}.
\label{eq37}
\end{equation}
\noindent
By integrating and returning to dimensional quantities in Eq.~(\ref{eq34}),
the result (\ref{eq37}) leads to
\begin{equation}
\dbT\ocF_{\rm TM}(a,T)=\frac{k_B b\vl\zeta(3)}{4\pi a^3\omega_p^2}T^3.
\label{eq38}
\end{equation}
\noindent
One can see that it is a correction of the higher order in $T$ than in
Eq.~(\ref{eq28}).

Now we turn our attention to the case $l=0$, $\alpha={\rm TE}$.
Using the dielectric permittivity (\ref{eq29}), the reflection coefficient
$\rTE(0,y,T)$ takes the form
\begin{equation}
\rTE(0,y,T)=-
\frac{\sqrt{1+\delta(T)y}-\sqrt{\delta(T)y}}{\sqrt{1+\delta(T)y}+\sqrt{\delta(T)y}},
\label{eq39}
\end{equation}
\noindent
where
\begin{equation}
\delta(T)=\frac{\bt T^2}{\tvt\tilde{\omega}_p^2}\ll 1.
\label{eq40}
\end{equation}

Expanding Eq.~(\ref{eq39}) up to the lowest order in $\delta(T)$, we have
\begin{equation}
\rTE(0,y,T)=-1+2\sqrt{\delta(T)y}.
\label{eq41}
\end{equation}
\noindent
Then from Eqs.~(\ref{eq39}) and (\ref{eq41}) one obtains
\begin{equation}
\rTE(0,y,0)=-1, \quad \dT\rTE(0,y,T)=2\sqrt{\delta(T)y}.
\label{eq42}
\end{equation}

Substituting Eq.~(\ref{eq42}) in Eq.~(\ref{eq32}) for $\alpha=$TE, we arrive at
\begin{eqnarray}
&&
\dbT\ocF_{\rm TE}(a,T)=\frac{k_BT}{4\pi a^2}\sqrt{\delta(T)}\int_{0}^{\infty}
\!\!dy\frac{y^{3/2}}{e^y-1}
\nonumber \\
&&\phantom{\dbT\ocF_{\rm TM}(a,T)}
=\frac{3k_B\sqrt{b}c\zeta(5/2)}{16\sqrt{2\pi}a^{5/2}\omega_p\sqrt{\vt}}T^2.
\label{eq43}
\end{eqnarray}
\noindent
One can see that although $\dbT\ocF_{\rm TE}$ is of the lower order than
$\dbT\ocF_{\rm TM}$ defined in Eq.~(\ref{eq38}), it is of the higher order in $T$
than $\daT\ocF$ defined in Eq.~(\ref{eq28}).

The sum of all terms with $l\geqslant 1$ in Eq.~(\ref{eq32}) results in the same
temperature dependence at low $T$ as in Eq.~(\ref{eq43}), i.e.,
\begin{equation}
\dcT\ocF_{\rm TM}(a,T)\sim \dcT\ocF_{\rm TE}(a,T)\sim T^2
\label{eq44}
\end{equation}
\noindent
(see Appendix A). Thus, by comparing Eqs.~(\ref{eq28}), (\ref{eq38}), (\ref{eq43}),
and (\ref{eq44}), we conclude that the leading low-temperature behavior of the
thermal correction to the Casimir energy between metallic plates with perfect
crystal lattices found within the Lifshitz theory using the alternative
Drude-like response functions is determined by the implicit term in Eq.~(\ref{eq18})
and takes the form
\begin{equation}
\dT\ocF(a,T)=-\frac{3c\zeta(3/2)\zeta(5/2)}{32\pi\omega_pa^{5/2}\sqrt{\vt\hbar}}
(k_BT)^{3/2}\!\!\!\! ~.
\label{eq45}
\end{equation}

{}From Eq.~(\ref{eq45}) it is easy to evaluate the dominant contribution to the
Casimir entropy at arbitrarily low temperature
\begin{eqnarray}
S(a,T)&=&-\frac{\partial \dT\ocF(a,T)}{\partial T}
\nonumber \\
&=&\frac{9k_Bc\zeta(3/2)\zeta(5/2)}{64\pi\omega_pa^{5/2}\sqrt{\vt\hbar}}
\sqrt{k_BT}.
\label{eq46}
\end{eqnarray}
\noindent
It is seen that the Casimir entropy is positive and goes to zero with
vanishing temperature as it should be in accordance to the third law of
thermodynamics, the Nernst heat theorem \cite{60}. This makes the
nonlocal Drude-like response function (\ref{eq6}) preferable as compared to
the conventional Drude response (\ref{eq7}). For metals with perfect
crystal lattices the latter leads to the negative Casimir entropy at
zero temperature \cite{26}
\begin{eqnarray}
&&
S_D(a,T)=-\frac{k_B\zeta(3)}{16\pi a^2}\left[1-4\kappa
+12\kappa^2-\ldots\right]<0,
\label{eq47}
\end{eqnarray}
\noindent
where $\kappa\equiv c/(\omega_p a)$. This entropy depends on the
parameters of a system, such as the separation
distance between the plates and the plasma frequency of a metal, and, thus,
violates the Nernst heat theorem \cite{60,61}.

\section{Low-temperature behavior of the Casimir free energy and entropy in the
presence of defects of a crystal lattice}

Crystal lattices of real metallic samples unavoidably contain some small
fractions of defects (e.g., atoms which are different from the native atoms
of the lattice, vacancies etc.). In this case, with decreasing temperature
the relaxation parameter $\gamma(T)$ reaches at $T=T_0$ some
minimum value $\gamma_0>0$ and remains unchanged at $T<T_0$ \cite{58}.
Thus, for typical Au samples $\gamma_0\approx 5.3\times 10^{10}~$rad/s.
As discussed in Sec.~1, this fact was used in the literature \cite{31,32,33}
in an attempt to reconcile the Lifshitz theory using the Drude model with the
Nernst heat theorem. It was shown, however, that with decreasing temperature
the Casimir entropy takes a negative constant value over the wide temperature
interval and abruptly jumps to zero only at $T<10^{-3}~$K which is somewhat
nonphysical \cite{20}.
Because of this, it is interesting to find the low-temperature behavior of
the Casimir entropy calculated using the nonlocal Drude-like response functions
for metals with defects of a crystal structure.

Below we consider the temperature interval $0\leqslant T<T_0$ where
$\gamma(T)=\gamma_0={\rm const}$. In this interval, the nonlocal Drude-like
permittivities are given by Eq.~(\ref{eq29}) where one should replace the term
$\bt T^2$ with $\gt=2a\gamma_0/c$. These permittivities do not possess an
explicit dependence on $T$:
$\teps_{D,l}^{\rm Tr}=\teTD(\ri\tau l,y)$ and
$\teps_{D,l}^{\rm L}=\teLD(\ri\tau l,y)$.
As a result, the thermal correction (\ref{eq10}) takes the same form as an
implicit thermal correction (\ref{eq20})
\begin{equation}
\dT\acF\aT=\frac{k_BT}{8\pi a^2}\left[\sum_{l=0}^{\infty}{\vphantom{\sum}}^{\prime}
\Phi_{\alpha}(\tau l)-\int_0^{\infty}\!\!\!dt\,\Phi_{\alpha}(\tau t)\right],
\label{eq48}
\end{equation}
\noindent
where now
\begin{equation}
\Phi_{\alpha}(x)\equiv \int_x^{\infty}\!\!\!ydy\ln\left[1-r_{\alpha}^2(\ri x,y)
e^{-y}\right].
\label{eq49}
\end{equation}
\noindent
The reflection coefficients $\rA(\ri x,y)$ are given by the right-hand sides
of Eq.~(\ref{eq24}) where
$\teTD{\vphantom{\teLD}}^{(0)}$, $\teLD{\vphantom{\teLD}}^{(0)}$
should be replaced with
$\teTD{\vphantom{\teLD}}_{\!,l}$, $\teLD{\vphantom{\teLD}}_{,l}$
specified above.

Using the Abel-Plana formula, Eq.~(\ref{eq48}) can be rewritten similar to
Eq.~(\ref{eq22})
\begin{equation}
\dT\acF\aT=\frac{\ri k_BT}{8\pi a^2}\int_0^{\infty}\!\!dt
\frac{\Phi_{\alpha}(\ri\tau t)-\Phi_{\alpha}(-\ri\tau t)}{e^{2\pi t}-1}
.
\label{eq50}
\end{equation}
\noindent
In order to obtain the behavior of $\dT\acF$ at low $T$, it is convenient
to expand $\Phi_{\alpha}$ up to the first power in $x$
\begin{equation}
\Phi_{\alpha}(x)=\Phi_{\alpha}(0)+x\Phi_{\alpha}^{\prime}(0),
\label{eq51}
\end{equation}
\noindent
where
\begin{equation}
\Phi_{\alpha}^{\prime}(0)=-2\int_{0}^{\infty}\!\!\!ydy
\frac{\rA(0,y)r_{\alpha}^{\prime}(0,y)e^{-y}}{1-r_{\alpha}^{2}(0,y)e^{-y}}
\label{eq52}
\end{equation}

We start with the transverse magnetic polarization $\alpha={\rm TM}$.
In this case, we use the parameter (\ref{eq34}) which becomes
temperature-independent because $\tilde{\gamma}(T)=\bt T^2$ is now equal to
$\tilde{\gamma_0}$
\begin{equation}
\beta_0=\frac{\gt\tvl}{\tilde{\omega}_p^2}\ll 1.
\label{eq53}
\end{equation}
\noindent
An extreme smallness of this parameter is evident because
$\gt/\tilde{\omega}_p\sim 10^{-6}$, $\tvl\sim 10^{-2}$,
whereas $\tilde{\omega}_p\sim 1$ at $a=10~$nm and
increases with increasing separation between the plates.

Expanding the reflection coefficient $\rTM(0,y)$ up to the first power
of small parameter (\ref{eq53}), we find
\begin{equation}
\rTM(0,y)=1-2\beta_0y.
\label{eq54}
\end{equation}

In a similar way,
\begin{equation}
r_{\rm TM}^{\prime}(0,y)=\left.
\frac{\partial\rTM(x,y)}{\partial x}\right|_{x=0}=-2\beta_0
\left(\frac{1}{\tvl}+\frac{y}{\gt}\right).
\label{eq55}
\end{equation}

Substituting Eqs.~(\ref{eq54})  and (\ref{eq55}) in Eq.~(\ref{eq52}), in the lowest order
of the small parameter $\beta_0$ one obtains
\begin{eqnarray}
\Phi_{\rm TM}^{\prime}(0)&=&4\beta_0\int_{0}^{\infty}\!\!\!ydy
\frac{(\tvl)^{-1}+\gt^{-1}y}{e^y-1}
\nonumber \\
&=&4\beta_0\left[\frac{\pi^2}{6\tvl}+\frac{2\zeta(3)}{\gt}\right].
\label{eq56}
\end{eqnarray}

Using Eqs.~(\ref{eq51})  and (\ref{eq56}), we have
\begin{equation}
\Phi_{\rm TM}(\ri\tau t)-\Phi_{\rm TM}(-\ri\tau t)=
8\ri\beta_0\tau t \left[\frac{\pi^2}{6\tvl}+\frac{2\zeta(3)}{\gt}\right].
\label{eq57}
\end{equation}

Then we substitute Eq.~(\ref{eq57})  in Eq.~(\ref{eq50}) and arrive at
\begin{eqnarray}
&&
\dT\ocF_{\rm TM}(a,T)=-\frac{k_BT}{\pi a^2}
\beta_0\tau\left[\frac{\pi^2}{6\tvl}+\frac{2\zeta(3)}{\gt}\right]
\nonumber \\
&&~~~~~~~~~~~~~~~~~~~~~~\times
\int_{0}^{\infty}\frac{tdt}{e^{2\pi t}-1}.
\label{eq58}
\end{eqnarray}

After integration and returning to the dimensional quantities, we finally find
\begin{equation}
\dT\ocF_{\rm TM}(a,T)=-\frac{(k_BT)^2}{12 a^2\hbar\omega_p^2}
\left[\gamma_0\frac{\pi^2}{6}+\frac{\vl}{a}\zeta(3)\right] .
\label{eq59}
\end{equation}

Now we continue with the transverse electric polarization, $\alpha={\rm TE}$.
The parameter (\ref{eq40}) used in this case in Sec.~III also becomes
temperature-independent
\begin{equation}
\delta_0=\frac{\gt}{\tvt\tilde{\omega}_p^2}\ll 1.
\label{eq60}
\end{equation}
\noindent
This parameter takes the maximum value $\approx 4\times 10^{-4}$ at $a=10~$nm
and further decreases with increasing separation between the plates.
The reflection coefficient $\rTE(0,y)$ takes the same form as in Eq.~(\ref{eq39})
\begin{equation}
\rTE(0,y)=-
\frac{\sqrt{1+\delta_0y}-\sqrt{\delta_0y}}{\sqrt{1+\delta_0y}+\sqrt{\delta_0y}}.
\label{eq61}
\end{equation}

Expanding in Eq.~(\ref{eq61}) up to the lowest order of the parameter (\ref{eq60}),
one obtains
\begin{equation}
\rTE(0,y)=-1+2\sqrt{\delta_0y}.
\label{eq62}
\end{equation}
\noindent
Then, by expanding the derivative of $\rTE(x,y)$ with respect to $x$ at
$x=0$, we have
\begin{equation}
r_{\rm TE}^{\prime}(0,y)=\frac{\sqrt{\delta_0}}{\gt}
\left(-\frac{\gt}{\tvt\sqrt{y}}+\sqrt{y}\right).
\label{eq63}
\end{equation}
\noindent
Substituting Eqs.~(\ref{eq62}) and (\ref{eq63}) in Eq.~(\ref{eq52}), in the
lowest order of the parameter (\ref{eq60}), one finds
\begin{eqnarray}
&&
\Phi_{\rm TE}^{\prime}(0)=2\frac{\sqrt{\delta_0}}{\gt}\left(
-\frac{\gt}{\tvt}\int_{0}^{\infty}\!\!\frac{\sqrt{y}dy}{e^y-1}+
\int_{0}^{\infty}\!\!\frac{y\sqrt{y}dy}{e^y-1}\right)
\nonumber \\
&&~~~~~~~~
=\frac{\sqrt{\pi\delta_0}}{\gt}\left[\frac{3}{2}
\,\zeta\!\left(\frac{5}{2}\right)-\frac{\gt}{\tvt}
\,\zeta\!\left(\frac{3}{2}\right)\right].
\label{eq64}
\end{eqnarray}
\noindent

Using Eq.~(\ref{eq64}), we have from Eq.~(\ref{eq51})
\begin{eqnarray}
&&
\Phi_{\rm TE}(\ri\tau t)-\Phi_{\rm TE}(-\ri\tau t)=2\ri\tau t
\frac{\sqrt{\pi\delta_0}}{\gt}
\nonumber \\
&&~~~~~~~\times
\left[\frac{3}{2}
\,\zeta\!\left(\frac{5}{2}\right)-\frac{\gt}{\tvt}
\,\zeta\!\left(\frac{3}{2}\right)\right]
\label{eq65}
\end{eqnarray}
\noindent
and substituting this in Eq.~(\ref{eq50}) arrive at
\begin{eqnarray}
&&
\dT\ocF_{\rm TE}(a,T)=-\frac{k_BT}{4\sqrt{\pi} a^2}
\frac{\tau\sqrt{\delta_0}}{\gt}
\nonumber \\
&&~~~~~\times
\left[\frac{3}{2}
\,\zeta\!\left(\frac{5}{2}\right)-\frac{\gt}{\tvt}
\,\zeta\!\left(\frac{3}{2}\right)\right]
\int_{0}^{\infty}\frac{tdt}{e^{2\pi t}-1}.
\label{eq66}
\end{eqnarray}
\noindent
Integrating in Eq.~(\ref{eq66}) and returning to the dimensional quantities, we finally
obtain
\begin{eqnarray}
&&
\dT\ocF_{\rm TE}(a,T)=-\frac{(k_BT)^2\sqrt{\pi}c}{48 a^{5/2}\hbar\omega_p
\sqrt{2\gamma_0\vt}}
\nonumber \\
&&~~~~~~~~~~~~~~~\times
\left[\frac{3}{2}
\,\zeta\!\left(\frac{5}{2}\right)-\frac{2a\gamma_0}{\vt}
\,\zeta\!\left(\frac{3}{2}\right)\right] .
\label{eq67}
\end{eqnarray}

As can be seen from Eqs.~(\ref{eq59}) and (\ref{eq67}), the dominant contribution to
the thermal correction is given by the TE mode due to the much larger coefficient near
the same power in $T$. This is ultimately determined
by the fact that $\beta_0\sim 10^{-8}$ whereas  $\delta_0\sim 10^{-4}$.
Because of this, the leading low-temperature behavior of the thermal
correction to the Casimir energy between metallic plates with typical
concentration of impurities found using the nonlocal Drude-like response
functions is given by
\begin{eqnarray}
&&
\dT\ocF(a,T)=-\frac{\sqrt{\pi}c}{48 a^{5/2}\hbar\omega_p
\sqrt{2\gamma_0\vt}}(k_BT)^2
\nonumber \\
&&~~~~~~~~~~~~~~~\times
\left[\frac{3}{2}
\,\zeta\!\left(\frac{5}{2}\right)-\frac{2a\gamma_0}{\vt}
\,\zeta\!\left(\frac{3}{2}\right)\right] .
\label{eq68}
\end{eqnarray}

The respective Casimir entropy at low temperature takes the form
\begin{eqnarray}
&&
S(a,T)=\frac{\sqrt{\pi}k_Bc}{24 a^{5/2}\hbar\omega_p
\sqrt{2\gamma_0\vt}}k_BT
\nonumber \\
&&~~~~~~~~~~~~~~~\times
\left[\frac{3}{2}
\,\zeta\!\left(\frac{5}{2}\right)-\frac{2a\gamma_0}{\vt}
\,\zeta\!\left(\frac{3}{2}\right)\right] .
\label{eq69}
\end{eqnarray}
\noindent
Thus, for metals with impurities the Casimir entropy calculated
using the alternative Drude-like response fun\-ctions monotonously
decreases to zero with vanishing temperature in accordance with
the Nernst heat theorem. Note also that the entropy (\ref{eq69})
remains positive at all separations below approximately $20~\mu$m,
i.e., in the region related to the Casimir effect. These properties
of Eq.~(\ref{eq69}) are advantageous in comparison with the conventional
Drude model. The latter leads to the negative Casimir entropy which
abruptly jumps to zero for metals with defects of a crystal lattice only
at very low temperature.

%%%%%%%%%%%%%%%%%%%%%%%%%%%%%%%%%%%%%
\section{Conclusions and discussion}

In the foregoing, we have elucidated thermodynamic properties
of the Casimir interaction calculated in the framework of the
Lifshitz theory using the spatially nonlocal Drude-like response
functions to quantum fluctuations. Although the Lifshitz theory
is a well-studied branch of thermal quantum field theory,
it suffers from a serious flaw known as the Casimir puzzle.
The problem lies in the fact that theory is in disagreement with
the measurement data and thermodynamic constraints
when using the conventional Drude response function to
quantum fluctuations. Recently it was shown, however,
that an agreement of the Lifshitz theory with the measurement
data can be restored when using alternative nonlocal
response functions which ensure almost the same response, as the
standard Drude model, to the on-shell fluctuations but respond
differently to quantum fluctuations off the mass shell
\cite{48}. However, the thermodynamic problem in the Lifshitz
theory remained unresolved.

According to the above results, the Lifshitz theory using the
suggested nonlocal Drude-like response functions is in complete
agreement with the requirements of thermodynamics. Using rigorous
analytic methods, it is shown that for metals with perfect crystal
lattices the Casimir entropy calculated using these functions is
positive and monotonously goes to zero as the square root of
temperature when the temperature vanishes. The same characteristic
properties are preserved for crystal lattices possessing some fraction
of defects with the only difference that the Casimir entropy
vanishes linearly in temperature. This means that the Lifshitz
theory combined with nonlocal Drude-like response
functions satisfies the third law of thermodynamics, the Nernst
heat theorem.

Taking into consideration that the suggested nonlocal response
functions to quantum fluctuations not only make the theoretical
predictions consistent with the measurement data, but also
secure an agreement between two fundamental theories, they
constitute a serious alternative to the commonly accepted
Drude model and deserve further investigation.

%%%%%%%%%%%%%%%%%%%%%%%%%%%%%%%%%%%%%%%%%%%
\section*{ACKNOWLEDGMENTS}

This work was supported by the Peter the Great Saint Petersburg Polytechnic
University in the framework of the Russian state assignment for basic research
(project N FSEG-2020-0024).
V. M. M.~was partially funded by the Russian Foundation for Basic
Research, grant number 19-02-00453 A.
V. M. M.\ was also partially supported by the Russian Government
Program of Competitive Growth of Kazan Federal University.

%%%%%%%%%%%%%%%%%%%%%%%%%%%%%%%%%%%%%%%%
\appendix
\section{Contribution from nonzero Matsubara %{\protect \\}
frequencies}
\setcounter{equation}{0}
\renewcommand{\theequation}{A\arabic{equation}}

Here, we derive low-temperature asymptotic behavior of the
contribution to explicit thermal correction (\ref{eq32}) from
all Matsubara terms with $l\geqslant 1$. For this purpose, in each term
of Eq.~(\ref{eq32})  with $l\geqslant 1$ we introduce the new integration
variable $z=y-\tau l$ and obtain
\begin{equation}
\dcT\acF(a,T)=-\frac{k_BT}{4\pi a^2}\sum_{l=1}^{\infty}e^{-\tau l}
X_{\alpha,l}(\tau),
\label{A1}
\end{equation}
\noindent
where
\begin{equation}
X_{\alpha,l}(\tau)
=\int_{0}^{\infty}\!\!\!dz(z+\tau l)
\frac{\rA(\ri\tau l,z+\tau l,0)\dT\rA(\ri\tau l,z+\tau l,\tau)}{e^z-
r_{\alpha}^2(\ri\tau l,z+\tau l,0)e^{-\tau l}}.
\label{A2}
\end{equation}

Note that in Eq.~(\ref{A2}) we have replaced an explicit dependence of
$\dT\rA$ on $T$ with a dependence on $\tau$ using the fact that
$T=\hbar c\tau/(4\pi ak_B)$. The thermal correction $\dT\rA$ on $T$
depends on $T$ as a parameter owing to the temperature-dependent
relaxation parameter. As discussed in Sec.~III, at temperatures below
the liquid helium temperature the dimensionless relaxation parameter
is of the order of $T^2$. Below we present it in the form
\begin{equation}
\tilde\gamma (T)=\frac{2a\gamma(T)}{c}=\bt T^2=\tilde{\bt}\tau^2,
\label{A3}
\end{equation}
\noindent
where
\begin{equation}
\bt=\frac{2ab}{c},\quad
\tilde{\bt}=\frac{c\hbar^2b}{8\pi^2k_B^2a}.
\label{A4}
\end{equation}

The reflection coefficients $\rA(\ri\tau l,z+\tau l,0)$ are given by Eq.~(\ref{eq24})
where $x$ should be replaced with $\tau l$ and $y$ with $z+\tau l$.
To obtain the coefficients $\rA(\ri\tau l,z+\tau l,\tau l)$ from Eq.~(\ref{eq24}),
one should also replace
$\teTD\vphantom{\teTD}^{(0)}$,  $\teLD\vphantom{\teLD}^{(0)}$ with
$\teTD\vphantom{\teTD}_{\!,l}$,  $\teLD\vphantom{\teLD}_{,l}$ defined
in Eq.~(\ref{eq29}) taking into account Eq.~(\ref{A3}).

Now we begin with the transverse magnetic polarization, $\alpha={\rm TM}$,
and expand both reflection coefficients  $\rTM(\ri\tau l,z+\tau l,0)$ and
$\rTM(\ri\tau l,z+\tau l,\tau)$ in the powers of small $\tau$
\begin{eqnarray}
&&
\rTM(\ri\tau l,z+\tau l,0)=1-\frac{2\tvl lz}{\tilde{\omega}_p^2}\tau
\nonumber \\
&&~
 +
\frac{2l^2}{\tilde{\omega}_p^4}\left[-\tilde{\omega}_p^2(1+\tvl)+
2(\tvl)^2z^2\right]\tau^2+O(\tau^{5/2}),
\label{A5} \\
&&
\rTM(\ri\tau l,z+\tau l,\tau)=1-\frac{2\tvl l z}{\tilde{\omega}_p^2}\tau
\nonumber \\
&&
-
\frac{2l^2}{\tilde{\omega}_p^4}\left[\tilde{\omega}_p^2\left(1+\tvl+
\frac{\tilde{\bt}\tvl}{l^2}z\right)-
2(\tvl)^2z^2\right]\tau^2+O(\tau^{5/2}).
\nonumber
\end{eqnarray}

Then from Eq.~(\ref{eq30}) one obtains
\begin{equation}
\dT\rTM(\ri\tau l,z+\tau l,\tau)=-\frac{2\tilde{\bt}\tvl z}{\tilde{\omega}_p^2}\tau^2.
\label{A6}
\end{equation}
\noindent
Substituting the first line of Eq.~(\ref{A5}) and Eq.~(\ref{A6}) in Eq.~(\ref{A2}),
we find in the lowest order
\begin{equation}
X_{{\rm TM},l}(\tau)=-\frac{2\tilde{\bt}\tvl}{\tilde{\omega}_p^2}\tau^2
\int_{0}^{\infty}\!\!\frac{z^2dz}{e^z-1}
=-\frac{4\tilde{\bt}\tvl}{\tilde{\omega}_p^2}\zeta(3)\tau^2.
\label{A7}
\end{equation}
\noindent
Then from Eq.~(\ref{A1}) one arrives at
\begin{equation}
\dcT\ocF_{\rm TM}(a,T)=
\frac{k_BT\tilde{\bt}\tvl\tau^2}{\pi a^2\tilde{\omega}_p^2}\zeta(3)
\sum_{l=1}^{\infty}e^{-\tau l}.
\label{A8}
\end{equation}
\noindent
Performing the summation in $l$ and returning to the dimensional quantities
with account of Eq.~(\ref{A4}), in the lowest order we finally find
\begin{eqnarray}
\dcT\ocF_{\rm TM}(a,T)&=&
\frac{k_BT\tilde{\bt}\tvl\tau^2}{\pi a^2\tilde{\omega}_p^2}
\frac{\zeta(3)}{e^{\tau}-1}
\nonumber \\
&=&
\frac{\hbar c b \vl\zeta(3)}{8\pi^2a^4\omega_p^2}T^2.
\label{A9}
\end{eqnarray}
\noindent
This result is in accordance with Eq.~(\ref{eq44}).

We continue  with the transverse electric polarization, $\alpha={\rm TE}$,
and expand the coefficients  $\rTE(\ri\tau l,z+\tau l,0)$ and
$\rTE(\ri\tau l,z+\tau l,\tau)$ in the powers of $\tau$
\begin{eqnarray}
&&
\rTE(\ri\tau l,z+\tau l,0)=-1+\frac{2\sqrt{lz}}{\tilde{\omega}_p\sqrt{\tvt}}
\sqrt{\tau}-\frac{2zl}{\tvt\tilde{\omega}_p^2}\tau
\nonumber \\
&&~
 +
\frac{l^{3/2}(\tvt\tilde{\omega}_p^2-\tilde{\omega}_p^2+
z^2)}{(\tvt)^{3/2}\tilde{\omega}_p^3\sqrt{z}}
\tau^{3/2}+O(\tau^{2}),
\label{A10} \\
&&
\rTE(\ri\tau l,z+\tau l,\tau)=-1+\frac{2\sqrt{lz}}{\tilde{\omega}_p\sqrt{\tvt}}
\sqrt{\tau}-\frac{2zl}{\tvt\tilde{\omega}_p^2}\tau
\nonumber \\
&&~
 +
\frac{l^{3/2}\left[\tvt\tilde{\omega}_p^2(1+\tilde{\bt}l^{-2}z)-
\tilde{\omega}_p^2+z^2\right]}{(\tvt)^{3/2}\tilde{\omega}_p^3\sqrt{z}}
\tau^{3/2}+O(\tau^{2}).
\nonumber
\end{eqnarray}

Using Eq.~(\ref{eq30}) for the thermal correction, we obtain
\begin{equation}
\dT\rTE(\ri\tau l,z+\tau l,\tau)=\frac{\tilde{\bt}\sqrt{z}}{\sqrt{\tvt}
\tilde{\omega}_p\sqrt{l}}\tau^{3/2}.
\label{A11}
\end{equation}
\noindent
Substituting Eq.~(\ref{A11}) and the first line of Eq.~(\ref{A10})  in
Eq.~(\ref{A2}),
one finds
\begin{eqnarray}
X_{{\rm TE},l}(\tau)&=&-\frac{\tilde{\bt}}{\sqrt{\tvt}
\tilde{\omega}_p\sqrt{l}}\tau^{3/2}
\int_0^{\infty}\!\!\frac{z^{3/2}}{e^z-1}dz
\nonumber \\
&=&-\frac{3\sqrt{\pi}\tilde{\bt}}{4\sqrt{\tvt}
\tilde{\omega}_p\sqrt{l}}\,\zeta\!\left(\frac{5}{2}\right)\tau^{3/2} .
\label{A12}
\end{eqnarray}
\noindent
Then, the substitution of Eq.~(\ref{A12}) in Eq.~(\ref{A1}) and summation in $l$
lead to
\begin{equation}
\dcT\ocF_{\rm TE}(a,T)=
\frac{3k_BT\tilde{\bt}\tau^{3/2}}{16\sqrt{\pi}a^2\sqrt{\tvt}
\tilde{\omega}_p}\,\zeta\!\left(\frac{5}{2}\right)
{\rm Li}_{1/2}\left(e^{-\tau}\right),
\label{A13}
\end{equation}
\noindent
where ${\rm Li}_{\beta}(z)$ is the polylogarithm function.

Taking into account that at small $\tau$ it holds \cite{62}
\begin{equation}
{\rm Li}_{1/2}\left(e^{-\tau}\right)
\approx\frac{\sqrt{\pi}}{\sqrt{\tau}},
\label{A14}
\end{equation}
\noindent
and returning to the dimensional quantities, we obtain
\begin{equation}
\dcT\ocF_{\rm TE}(a,T)=
\frac{3\hbar c^{3/2}b\zeta(5/2)}{64\pi a^3\sqrt{\tvt}
\tilde{\omega}_p}T^2,
\label{A15}
\end{equation}
\noindent
i.e., Eq.~(\ref{eq44}) is finally proven.

%%%%%%%%%%%%%%%%%%%%%%%%%%%%

%%%%%%%%%
\end{document}